# Disentangling photodoping, photoconductivity, and photosuperconductivity in the cuprates

R. El Hage[1], D. Sánchez-Manzano[1], V. Humbert[1], S. Carreira[1], V. Rouco[1], A Sander[1], F. Cuellar[2], K. Seurre[1], A. Lagarrigue[1], J. Briatico[1], J. Trastoy[1], J. Santamaría[2] and Javier E. Villegas[1,*]

[1]*Unité Mixte de Physique, CNRS, Thales, Université Paris Saclay, 91767 Palaiseau, France*

[2]*GFMC, Dpto. Física de Materiales. Universidad de Ciencias Físicas, Universidad Complutense de Madrid, 28040 Madrid, Spain*

The normal-state conductivity and superconducting critical temperature of oxygen-deficient $YBa_2Cu_3O_{7-\delta}$ can be persistently enhanced by illumination. Strongly debated for years, the origin of those effects –termed persistent photoconductivity (PPC) and photosuperconductivity (PPS)– has remained an unsolved critical problem, whose comprehension may provide key insights to harness the origin of high-temperature superconductivity itself. Here we make essential steps toward understanding PPS. While the models proposed so far assume that it is caused by a carrier-density increase (photodoping) observed concomitantly, our experiments contradict such conventional belief: we demonstrate that it is instead linked to a photo-induced decrease of the electronic scattering rate. Furthermore, we find that the latter effect and photodoping are completely disconnected and originate from different microscopic mechanisms since they present different wavelength and oxygen-content dependences as well as strikingly different relaxation dynamics. Besides helping disentangle photodoping, PPC, and PPS, our results provide new evidence for the intimate relation between critical temperature and scattering rate, a key ingredient in modern theories on high-temperature superconductivity.

[*]javier.villegas@cnrs-thales.fr

The transport properties of underdoped YBa$_2$Cu$_3$O$_x$ (YBCO) ($6.3 < x < 6.9$) can be altered via illumination with visible-UV light [1–14]. Following photoexcitation, and once in the dark, a decrease of the resistivity and an increase of the superconducting critical temperature T$_C$ are observed. These effects, termed persistent photoconductivity (PPC) and photosuperconductivity (PPS), are accompanied by electronic changes (a decrease in the Hall coefficient and an increase in Hall mobility) [3,8,14], an enhancement of the superconducting coherence lengths [10], as well as by a contraction of the *c* axis parameter [9]. These photoexcited states relax only at a high temperature (typically above 250K) [4,5,13]. Because PPC and PPS become stronger with decreasing *x* and vanish in samples with optimum oxygen stoichiometry, it is widely accepted that they are linked to the presence of oxygen vacancies, particularly in the basal CuO chains [9,15].

PPC and PPS attracted a great deal of attention since their discovery, in the early times of high-T$_C$ superconductivity. Their understanding is crucial for various reasons. First, explaining PPS may contribute to understanding the crux of the matter –the origin of high-T$_C$ superconductivity itself. Second, an analogous photoinduced switching of ground-states has been recently observed in other strongly correlated oxides, for example, manganites [16,17], iridates [18], nickelates [19,20] and LaAlO$_3$/SrTIO$_3$ heterostructures [21], which are also far from being fully understood. Since these systems share many electronic properties and structural features with cuprates, it is likely they also share some of the unknown mechanisms underlying their photosensitivity.

Despite much work, a clear picture of the microscopic mechanisms of PPS has not emerged. Because PPC and PPS are accompanied by an increase in the Hall number $n_H \equiv \left( d\rho_{xy}/dB \big|_{B=0} e \right)^{-1}$ (with $\rho_{xy}$ the Hall resistivity, $B$ the applied field, and $e$ the electron charge), existing models generally consider they both primarily originate from an enhancement of the carrier concentration (photodoping) whose microscopic origin was largely debated (see [1] for a review). Kudinov et al. [5] first proposed that photoexcitation produces electron-hole pairs which, followed by electron trapping and the promotion of holes to the conduction band, leads to the photodoping that ultimately enhances the conductivity and $T_C$. Electrons would be trapped by charged defects, such as oxygen vacancies or by CuO chain fragments [6]. In this scenario, the photoinduced effects relax as $T$ increases as electrons are released from their traps, and recombine. An alternative scenario was proposed by Osquiguil et al. [7] based on the idea that photoexcitation promotes the reordering of oxygen vacancies within the CuO chains, lengthening the average chain fragment, which expectedly produces an increase of carrier density in $CuO_2$ planes [22] and thus the enhancement of the YBCO superconducting properties.

While hybrid scenarios have also been proposed [23,24], in all cases, PPS is primarily attributed to photodoping. Contrarily, it is generally assumed that the concomitant changes in the Hall mobility $\mu_B$ [14] have no causal effect on the enhancement of superconducting properties. Yet, changes in $\mu_B$ reflect changes in the electron scattering [25,26], which has strong ties with the superconducting pairing in the cuprates in various ways. For example, it is well known that scattering by impurities [26] or disorder [27,28] depresses $T_C$ via pair breaking. Another example is the observation of quantum criticality –a scaling between $T_C$ and the T-linear scattering rate [29], which furthermore exhibits the same anisotropy as the d-wave gap [30,31]– pointing to a common origin for anisotropic scattering and pairing [32]. Because illumination modifies both $\mu_B$ and $T_C$, the question is whether a causal relationship exists and whether the mechanisms behind PPS go beyond the photodoping scenario.

The present experiments show that PPS is indeed strongly correlated with the photoinduced enhancement of $\mu_B$, but not with the variations of $n_H$. This finding challenges the idea that PPS is caused by photodoping and suggests instead that it is related to a change in the scattering rate. Furthermore, not only are the variations of the $\mu_B$ and $n_H$ decorrelated, but they also relax over very different time and temperature scales, implying that they originate from different microscopic mechanisms. Our observations allow for reconciling the main scenarios for the microscopic mechanisms by assigning separate roles to each of them. We propose that photodoping can be explained within the photocarrier excitation/electron trapping model, which would not play a major role in PPS, while oxygen ordering explains the photoinduced decrease of the scattering rate, and would more substantially contribute to PPS.

In our experiments, we compare oxygen-deficient c-axis $YBa_2Cu_3O_x$ (YBCO) films with optimally doped ones in which disorder has been induced by ion irradiation. All the films were grown on (001) $SrTiO_3$. Oxygen-deficient films (~30 nm thick) were grown by sputtering in a pure oxygen (3.4 mbar) atmosphere at T= 900º C. Samples with a nominal oxygen content $6.3 < x < 6.9$ were prepared by adjusting the pressure during sample cooldown following different lines of the pressure-temperature phase diagram [33,34]. Two samples were prepared for irradiation, the one (~ 30 nm thick) by sputtering as described above and the second (~50 nm thick) by pulsed laser deposition (λ=248 nm, energy ~1 J.cm$^{-2}$, repetition rate 5Hz) with oxygen pressure 0.36 mbar and T= 700°C. In the PLD, the oxygen pressure was raised to 800 mbar right after deposition and kept constant during cool-down to obtain nearly optimally doped films ($T_C$~85K). Each of these samples was diced in four pieces and subsequently irradiated with $O^+$ (E=110 keV) at different doses $10^{13} < d < 10^{14}$ ions cm$^{-2}$. As discussed elsewhere [35–37], in these conditions the $O^+$ track length across YBCO is ~150 nm. Thus, ions finish their course in the substrate and are not implanted within YBCO, where they create point defects (vacancies/interstitials) due to the displacement of atoms from the structure (mainly oxygen), at a ratio (Displacements Per Atom in the structure) $DPA = d \cdot 3.82 \cdot 10^{-16}$ cm$^{-2}$, as calculated using SRIM simulations [38,39]. X-ray diffraction confirms the quality of the films, shows the expected dependence on $x$ and $d$ [40,41], and demonstrates that the structural coherence is not affected by oxygen depletion but diminishes as $d$ increases (see Supplemental Material).

Photolithography was used to define a Hall bar for longitudinal $\rho_{xx}$ and Hall $\rho_{xy}$ resistivity measurements, carried out by injecting current and sensing voltages with a nanovoltmeter. These experiments were performed in a closed-cycle refrigerator equipped with an electromagnet to apply a magnetic field $B$, and an optical window for illumination. Three different light sources were used: two monochromatic LEDs [green $\lambda = 565 nm$ and near-UV $\lambda = 365 nm$), and a Xe arc source coupled to a filter to shine a deeper UV region of the spectrum ($\lambda = [260 - 320]\, nm$). From the transport measurements we calculate $n_H$ and $\mu_H \equiv \left(d\rho_{xy}/dB\big|_{B=0}\right)/\rho_{xx}$.

Fig. 1 summarizes the native properties of the films. (a) and (b) respectively show $\rho_{xx}(T)$ of films with different oxygen content $x$ (red data) and dose $d$ (blue). For oxygen-deficient samples, $\rho_{xx}$ decreases and $T_C$ (defined from the maximum of the derivative of $\rho_{xx}(T)$ as shown in Fig. S5) decreases as $x$ is decreased from the optimal $x$=6.9, until an insulating state sets for $x$< 6.3. That is detailed in Fig 1(c) and (d), where $T_C(x)$ displays a characteristic [34] plateau for 6.4 < $x$ < 6.6 and a fast drop for $x$ < 6.4. Supplemental Fig. S5 further supports that our samples present the characteristics of high-quality oxygen-deficient YBCO [42,43]. Irradiated samples show a distinct behavior, with a rapid $T_C$ decrease as $d$ is increased until an insulating state is obtained for $d = 2\,10^{14}$ ions cm$^{-2}$. Here superconductivity depression is not due to changes in the oxygen content, but to irradiation-induced structural disorder that enhances electronic scattering and pair-breaking [27]. We stress that no relationship exists between $x$ in the oxygen-deficient series and $d$ in the irradiated one: data are displayed in the same graphs for compacity, and to illustrate that $\rho_{xx}$ and $T_C$ span over the same range for both sets of samples, which calls for a comparison of their response to illumination.

Fig. 2 summarizes the illumination experiments, carried out always with the same protocol: first, we measured $\rho_{xx}(T)$ and the $\rho_{xy}(B)$ at T= 95 K in the dark. Following this, we illuminated the sample (optical power $P \sim 100.$ mW cm$^{-2}$) for 4 hours, turned off the light, and waited for 30 minutes in the dark to allow for thermalization. Then $\rho_{xy}(B)$ at T= 95K and $\rho_{xx}(T)$ (temperature ramped up from 3 K) were subsequently measured. Fig. 2 (a) shows the typical behavior of oxygen-deficient films. The main panel depicts $\rho_{xx}(T)$ before and after illumination, which induces a persistent increase of $T_c$ (PPS) and a decrease in the normal-state resistivity (PPC). The inset shows the corresponding Hall resistance measurements. Linear $\rho_{xy}(B)$ are measured whose slope corresponds the Hall coefficient $R_H$, allowing us to write $n_H = (R_H e)^{-1}$ and $\mu_H = R_H/\rho_{xx}$. One can see that illumination leads to a decrease in the Hall coefficient, that is, to an increase in $n_H$. While the free-electron model does not apply to the cuprates, it has been shown [14,25,44] that the variation $\Delta n_H$ at a fixed temperature corresponds to the carrier density change. Therefore, the Hall measurements indicate that illumination produces a carrier density increase (photodoping), as expected [1–14].

The photo-induced variations $\Delta T_c$, $\Delta n_H$ and $\Delta \mu_H$ (at T=95 K) are respectively displayed in Fig. 2(b), (c) and (d) as a function of $x$ (red symbols, oxygen-depleted samples) and of $d$ (blue symbols, irradiated samples), for various wavelengths (see legend).

For the oxygen-deficient set, Fig. 2 (c) shows that illumination produces a carrier density increase $\Delta n_H$ (photodoping) that is similar for all $x$ (within the error bars) and independent of the wavelength. In stark contrast, both $\Delta T_c$ and $\Delta \mu_H$ [Figs. 2 (b) and (d)] depend on the oxygen content and on the wavelength: $\Delta T_c$ and $\Delta \mu_H$ increase as the nominal oxygen content $x$ is reduced, and deep UV light produces larger $\Delta \mu_H$ (and correspondingly larger $\Delta T_c$) than green and near-UV light. The inescapable conclusion is that, while illumination causes a carrier concentration enhancement for all samples, there is no clear correlation between its magnitude (similar for all oxygen contents and wavelengths) and $\Delta T_c$. Instead, a strong correlation is observed between $\Delta T_c$ and $\Delta \mu_H$: both scale and show the same dependence on the oxygen content and the wavelength. This conclusion is unambiguously illustrated by $\Delta T_c$ vs. $\Delta n_H$ and $\Delta T_c$ vs. $\Delta\mu_H$ , respectively displayed in the insets of Figs. 2 (c) and (d): while data is scattered in the former case, a strong correlation is observed in the latter.

For the set irradiated samples (blue data), illumination produces a dose-independent $\Delta n_H$ [Fig 2 (c)] that, strikingly, is comparable to that of the oxygen-depleted samples. Thus, photodoping is not an exclusive property of oxygen-depleted samples. However, the irradiated samples show no photoinduced increase of $\mu_H$ nor $T_c$ [Fig 2 (b) and (d)].

The above findings suggest that photodoping, on the one hand, and photoinduced $\mu_H$ enhancement and the associated PPS on the other, are different effects arising from different microscopic mechanisms. This is further evidenced in what follows.

Figure 3 shows the typical temperature-dependent relaxation of the photo-induced effects for oxygen-depleted samples. For these experiments, after illumination at T = 95K, the temperature was ramped (~ 5K/min) to the target T and stabilized, and then the longitudinal and Hall resistivity were measured in the dark as a function of time $t$. Fig. 3 (a) and (b) show $n_H(t)$ and $\mu_H(t)$ across the relaxation, for several temperatures 250K < T < 295K. $n_H(t)$ and $\mu_H(t)$ relax towards the pristine levels displayed on the right side of the curves (data points were measured before illumination). Up to 260 K, the photodoping [Fig, 3 (a)] is persistent: $n_H$ stays above the pristine state level and presents no measurable change over 200 minutes. For higher temperatures, a relaxation is observed, at a rate that increases with increasing temperature [see also inset of Fig 3 (a)]. At room temperature. although relaxation is not complete over the experimental time, the carrier density enhancement has almost vanished after 200 minutes. $\mu_H(t)$ displays a strikingly different behavior [Fig. 3 (b)]. Already at the lowest T= 250 K, $\mu_H(t)$ decays almost linearly over time. Once the temperature increases, the relaxation rate increases much faster than for $n_H$: at T=280 K, $\mu_H$ nearly reaches the pristine state level within the experimental window, and at room temperature, the dynamic is inverted (in agreement with [14,24]) and completed in only a few minutes. This contrasts with the much longer relaxation time (> 200 minutes at room temperature) of the carrier density enhancement [Fig. 3 (a) and its inset]. The fact that $\Delta n_H$ and $\Delta \mu_H$ relax differently clearly evidences that they originate from different microscopic mechanisms. Interestingly, the observations discussed so far allow us to disentangle photodoping and PPS –generally supposed to be the two sides of the same coin [1]– and provide crucial clues for understanding their microscopic origin.

First, our experiments demonstrate that the photoinduced $\Delta n_H$ and $\Delta \mu_H$ are two uncorrelated phenomena. $\Delta n_H$ is similar for all *x*, illumination wavelengths, and *d*. Contrarily, $\Delta \mu_H$ strongly and monotonically depends on all those parameters: it is larger the lower the *x*, the most pronounced with deep-UV light, and non-existing in the case of ion-irradiated samples. Notice also that the relative mobility variation $\Delta \mu_H / \mu_H (pristine)$ ~ 30%-50 % is large (particularly in the low oxygen-content samples that show the largest PPC and PPS) despite a relatively modest $\Delta n_H / n_H (pristine)$ < 5 %. Furthermore, and most strikingly, the very different relaxation of $\Delta n_H$ and $\Delta \mu_H$ univocally demonstrates that they stem from different microscopic mechanisms. This is consistent with the earlier observation that, during illumination, $\mu_H$ and $n_H$ evolve differently [14,24]. It is therefore clear that the $\Delta n_H$ and $\Delta \mu_H$ are concomitant but essentially unrelated illumination effects.

Second, and crucially, our experiments show that the photoinduced $\Delta T_C$ is correlated with $\Delta \mu_H$, but not with $\Delta n_H$ Clear-cut evidence is provided by i) the oxygen-deficient samples, for which $\Delta T_C$ and $\Delta \mu_H$ are proportional and show the same dependence on the oxygen content and the illumination wavelength; and ii) the ion-irradiated samples in which, despite $\Delta n_H$ being similar that in oxygen-deficient ones, no $\Delta \mu_H$ and consequently no $\Delta T_C$ are observed.

The above conclusions call for an interpretation in terms of two different microscopic mechanisms, the one leading to $\Delta n_H$ and the other to $\Delta \mu_H$ and $\Delta T_C$.

Regarding $\Delta \mu_H$, it is helpful to recall Anderson's theory [25] that explains Hall effects in the cuprates [26] based on the relaxation rate of the quasiparticle-like excitations (spinons) in the interacting-electron (Luttinger) liquid. In this framework the Hall angle $\tan\theta_H \equiv \rho_{xy}/\rho_{xx}=\mu_H B$, reflects i) spinon-spinon scattering and ii) impurity (or defect) scattering [25]. While (i) depends on the doping level, (ii) does not [25,26]. If we also consider the fact that photodoping and mobility (Hall angle) variations are decorrelated, we conclude that the $\mu_H$ enhancement reflects a decrease in the impurity scattering rate.

One needs to explain now at a microscopic level why illumination reduces the impurity scattering rate. If we recall the two scenarios debated, namely i) photoexcitation plus electron trapping and ii) photoinduced oxygen ordering within the CuO chains, we find that the latter naturally explains the decrease in the impurity scattering if we consider that oxygen ordering reduces the impurity potentials introduced by oxygen vacancies. Furthermore, this also allows understanding of the associated $T_C$ increase, as it is well-known that in d-wave superconductors lower impurity scattering implies weaker pair-breaking [28] and thus higher $T_C$. Within this scenario, the presence of PPS in oxygen-depleted samples and its absence in ion-irradiated ones can be understood as follows. In oxygen-depleted YBCO, vacancies (and the related disorder) are essentially found within the CuO chains [1]. Here, the barrier for ion displacement and more specifically for oxygen jump between chain fragments is relatively low (~1 eV) [2], allowing for optical activation. However, in irradiated samples, vacancy-interstitial pairs are created for all atomic species. Furthermore, these defects, particularly oxygen ones (44% of the total according to SRIM simulations [39]), are not solely located in the CuO chains, but also in the $CuO_2$ planes, where the activation energy for ion displacement is up to ten times higher [28] than in the chains, thus precluding re-ordering by optical activation. Yet, defects in the $CuO_2$ planes play the dominant role in the large depression of carrier mobility and $T_c$ produced upon ion irradiation [27,28]. In conclusion, the absence of mobility enhancement and PPS in ion-irradiated samples results from the fact that illumination cannot "heal" the broader disorder that causes the superconductivity depression in them.

If oxygen ordering can explain the $\mu_H$ and $T_C$ enhancements, it cannot account for the concomitant photodoping because it is uncorrelated to those effects, as demonstrated by all our experimental observations. However, photodoping can be understood independently, in terms of photoexcitation of electrons that are captured either by charged defects (such as oxygen vacancies) [14,15] or by unoccupied p-levels of the oxygen atoms at the CuO chains [5,6], leaving excess holes in the $CuO_2$ planes. The fact that the observed carrier density variation is similar in all samples indicates that the density of "electron traps" is always large enough so that, within the chosen illumination time, there is no saturation. Indeed, this experimental condition has allowed us to show that photodoping is not the main ingredient for PPS.

In summary, we have studied oxygen-deficient and ion-irradiated YBCO films and evidenced that, contrary to conventional belief, photodoping and PPS have no causal relationship. We find instead that PPS is observed only in YBCO films in which illumination produces a significant carrier mobility enhancement, and particularly that the size of PPS scales with the photoinduced mobility enhancement. This unearths an interesting facet of the link between scattering and pairing in the superconducting cuprates [32], i.e. that it is controllable by light. These results call for examining the effect of VIS-UV light on other phenomena that emerge in the underdoped region of the cuprate's phase diagram. For example, charge density wave (CDW) order [45–48] whose competition with superconductivity can be tipped by disorder and scattering [49–51] as it has been shown in the case of external perturbations such as large magnetic fields [52] or strain [53,54]. These effects could be studied using techniques such as resonant inelastic X-ray Scattering [55,56] after illumination. Another worthwhile avenue of research stemming from the present study is looking into the possible universality of the microscopic mechanisms underlying light effects, in systems showing similar quantum critical transport phenomena [16–21].

**Acknowledgments**

Work supported by ERC grant N° 647100 "SUSPINTRONICS", French ANR grant ANR-17-CE30-0018-04 "OPTOFLUXONICS", COST action "Nanocohybri", Spanish AEI grants PID2020-118078RB-I00 and PCI2020-112093, and Madrid Regional Government grant Y2020/NMT-6661. JS thanks a Scholarship program Alembert funded by the IDEX Paris-Saclay, ANR-11-IDEX-0003-02. We (JS, J-E V) acknowledge funding from Flag ERA ERA-NET To2Dox project. We thank Y. Legall for ion irradiation at ICube (Strasbourg).

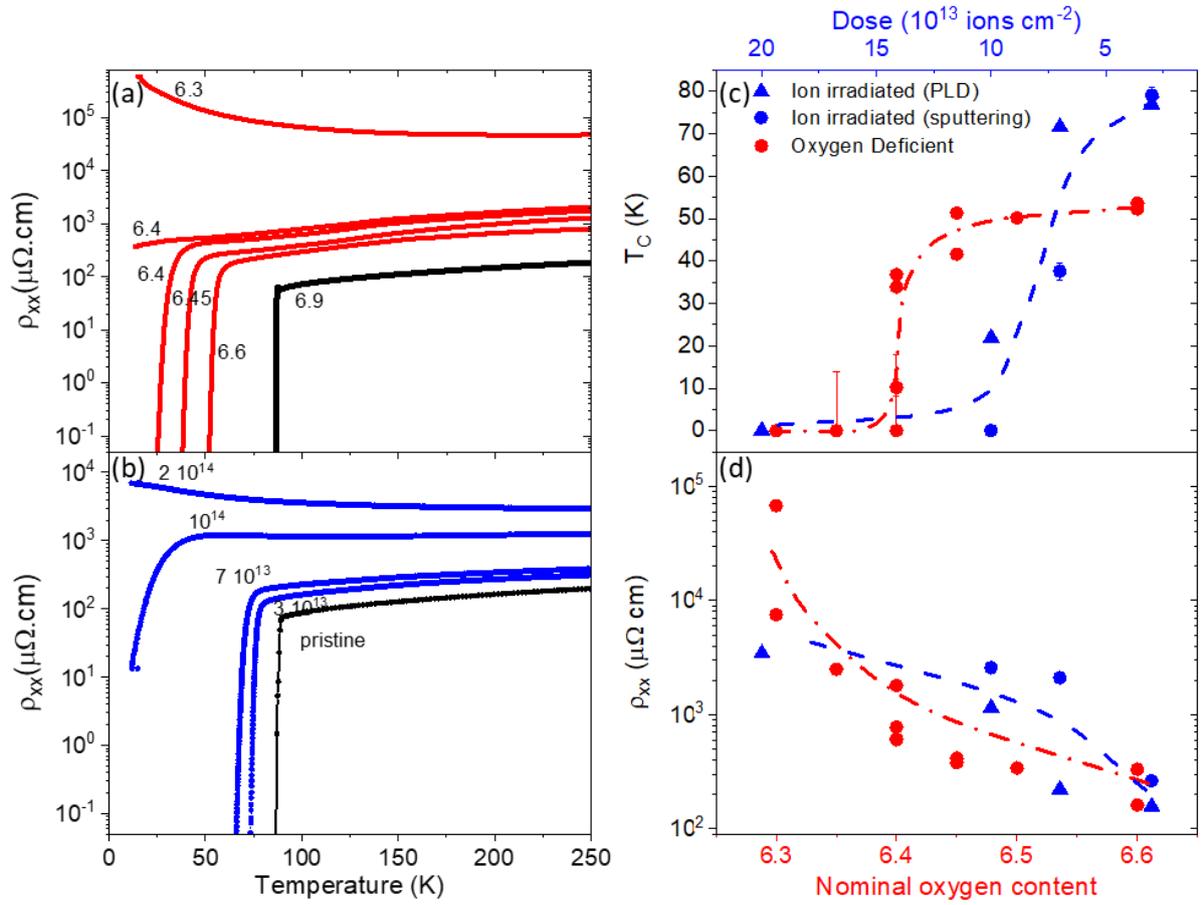

**Figure 1:** Temperature-dependent resistivity for (a) oxygen-deficient ($6.3 < x < 6.9$) and (b) ion-irradiated ($3\ 10^{13} < d < 2\ 10^{14}$ ion cm$^{-2}$) samples. (c) $T_C$ and (d) $\rho_{xx}(95\ K)$ vs. oxygen content $x$ (red data) and irradiation dose (blue data).

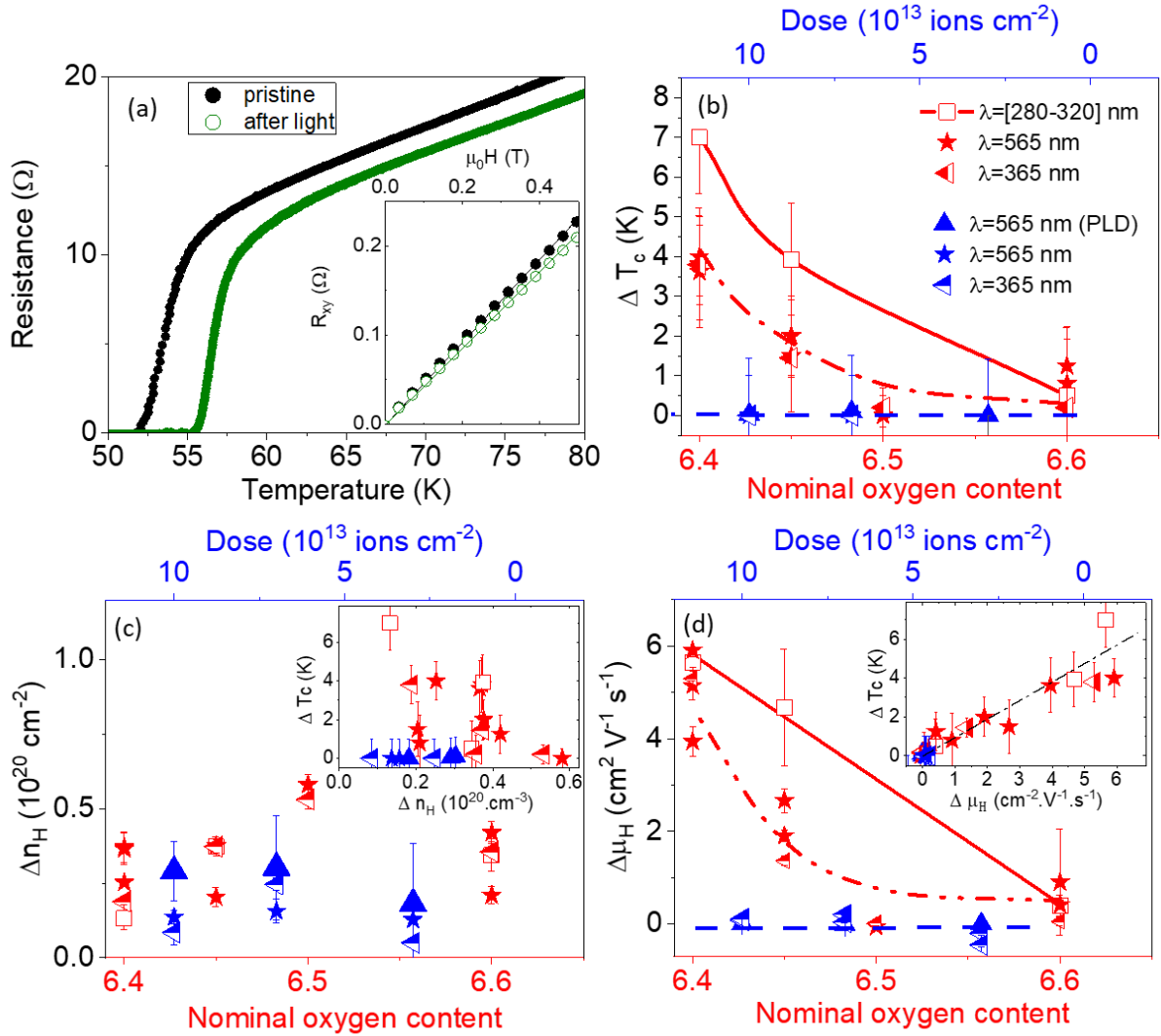

**Figure 2**: (a) Temperature-dependent resistivity before and after illumination, for an oxygen-deficient film ($x = 6.4$). The inset shows the corresponding Hall resistance at $T = 95K$. (b) Photoinduced changes $\Delta T_c$ (c) Hall number $\Delta n_H$ and (d) Hall mobility $\Delta \mu_H$ as a function of the nominal oxygen content (lower axis, red) and irradiation dose (upper x-axis), for different wavelengths (see legend). The irradiated set includes sputtering and PLD-grown samples (see legend), both types of samples behaving similarly. The insets in (c) and (d) show the $\Delta T_c$ vs. $\Delta n_H$ and $\Delta T_c$ vs. $\Delta \mu_H$ respectively. Notice that the latter shows a strong correlation. All the lines are guides to the eye.

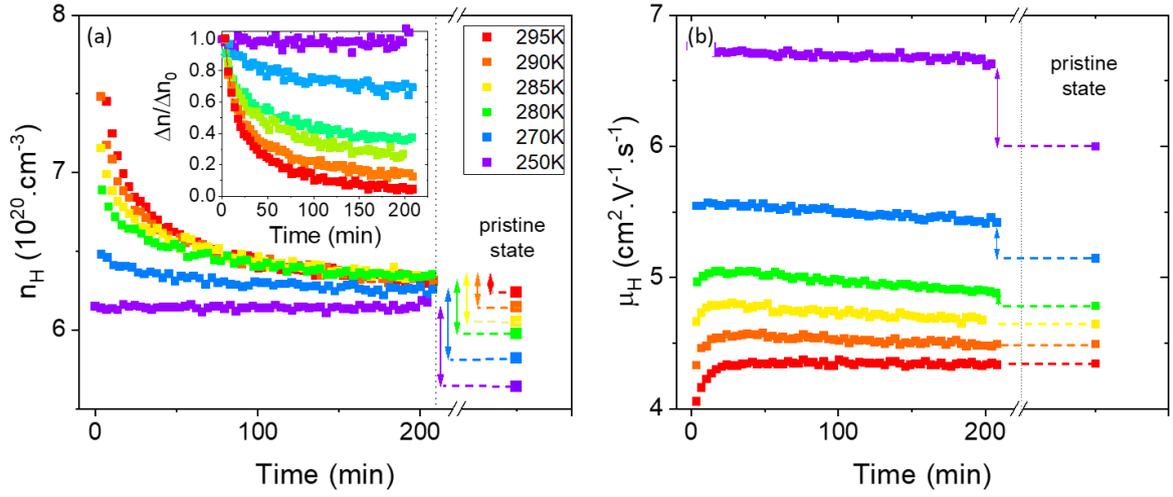

**Figure 3:** Time-dependent Hall number (a) and Hall mobility (b) measured in the dark at different temperatures (see legend), following illumination at T = 95K with $\lambda = 565\,nm$ and $P = 100\,mW.cm^{-2}$ (for a sample $x$=6.4). The reference levels on the right were measured before illumination. The double-sided arrows highlight the remnant enhancement at t=200 min. The inset in (a) shows the time-dependent relative variation of the carrier density $\Delta n/\Delta n_0$ where $\Delta n_0 = n_{t=0} - n_{pristine}$.

R. El Hage[1], D. Sánchez-Manzano[1], V. Humbert[1], S. Carreira[1], V. Rouco[1], A Sander[1], F. Cuellar[2], K. Seurre[1], A. Lagarrigue[1], J. Briatico[1], J. Trastoy[1], J. Santamaría[2] and Javier E. Villegas[1,*]

[1]*Unité Mixte de Physique, CNRS, Thales, Université Paris Saclay, 91767 Palaiseau, France*

[2]*GFMC, Dpto. Física de Materiales. Universidad de Ciencias Físicas, Universidad Complutense de Madrid, 28040 Madrid, Spain*

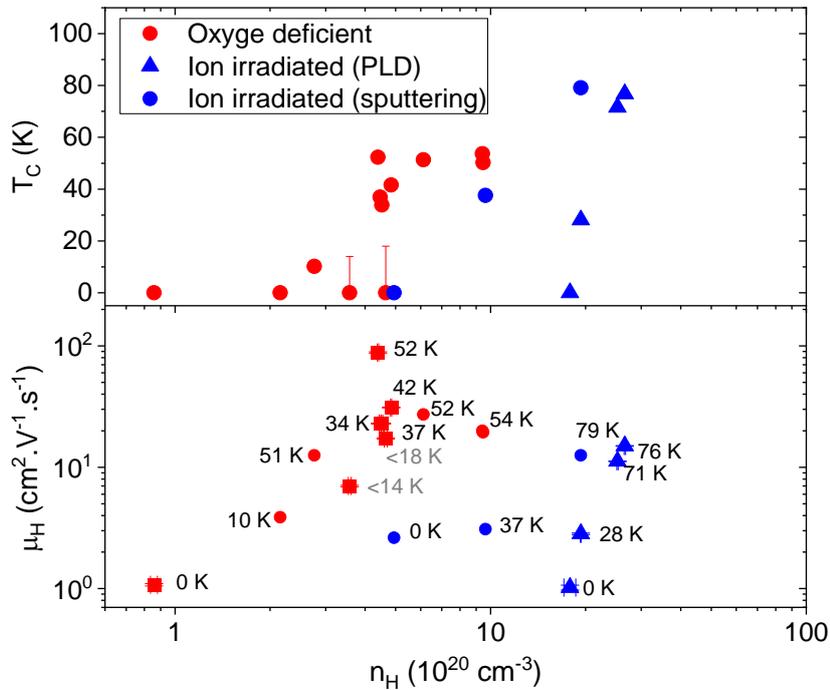

**Fig. S1** $T_C$ **(up)** and Hall mobility **(down)** *vs*- Hall number for the two sets of samples. Grey critical temperature labels indicate upper limits of the measurements. Non-superconducting samples are not discussed in the main text.

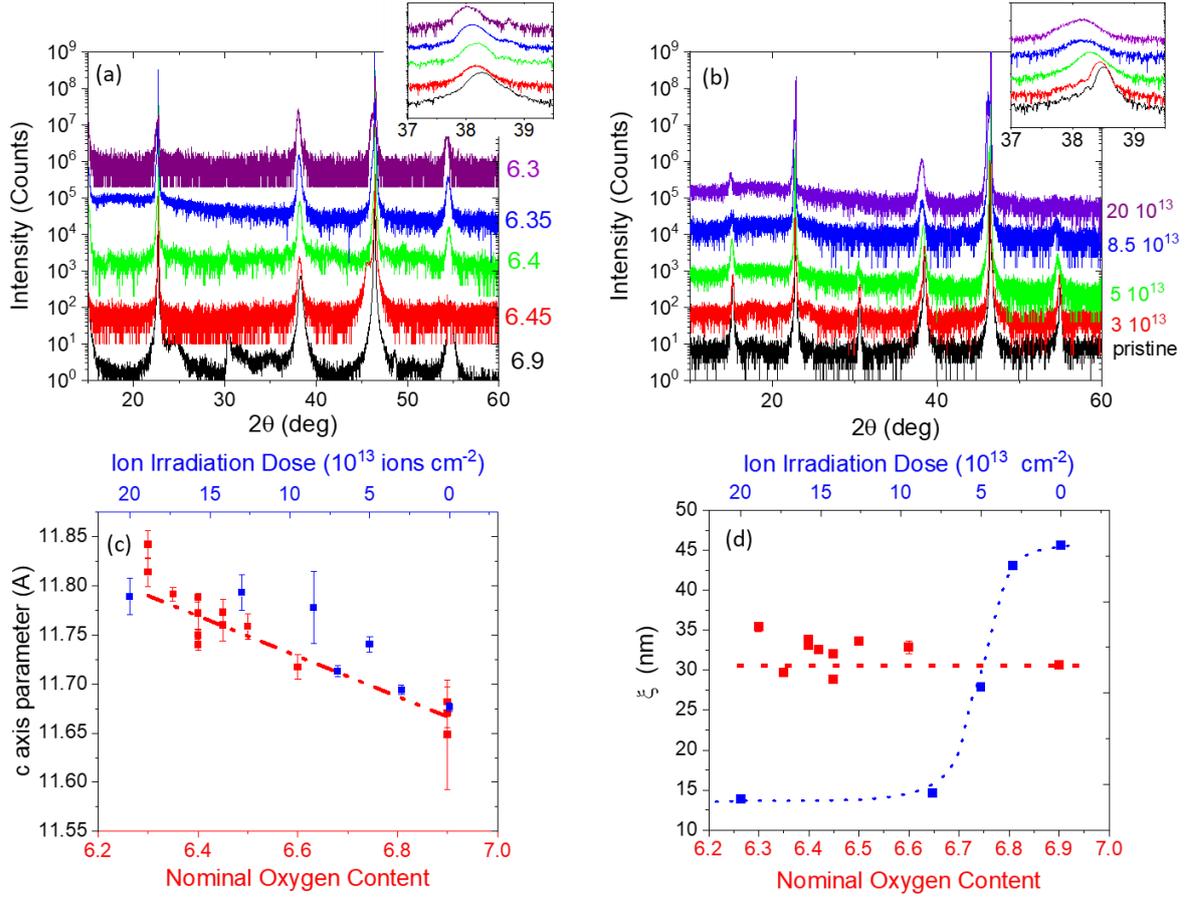

**Fig. S2** X-ray diffraction (scattering vector along c-axis) of **(a)** oxygen-deficient and **(b)** ion-irradiated YBCO films. Labels indicate oxygen-content $x$ and irradiation dose (in ions cm$^{-2}$). The YBCO (005) peak centered around $2\theta \approx 38°$ is shown in the inset to highlight the effect of both oxygen deficiency and ion irradiation on the YBCO structural properties. The spectra have been vertically shifted for clarity. **(c)** c-axis parameter as a function of the nominal oxygen content (red) and ion irradiation dose (blue). Calculated from the XRD spectra using Bragg law $2d_{hkl}sin\theta = n\lambda$ and assuming orthorhombic structure. As expected, the $c$ axis parameter increases as the oxygen content $x$ decreases [1] and also as the irradiation dose increases [2]. The straight line is the best fit for the oxygen-depleted series, which yields $c = A + Bx$, with $A$=13.1±0.4 Å and B=-0.20±0.1 Å. These parameters are comparable to those found in the literature $A = 12.85 \pm 0.4$ Å and $B = -0.171$, see e.g. [3]. From the data scattering around the linear regression, we can estimate the uncertainty of the nominal oxygen content $\Delta x \sim 0.1$ for the samples where photoinduced effects are studied $6.4 \leq x \leq 6.6$. (d) Structural correlation length along the c-axis as obtained from the Scherrer formula $\xi \approx 0.9\lambda/\Gamma cos\theta$, with $\theta$ the diffraction angle, $\lambda \sim 1.54$ Å the X-ray wavelength, and $\Gamma$ the 005 peak full-width at half-maximum. In the case of oxygen-depleted samples (red), the structural coherence is independent of the oxygen content and is limited by the sample thickness ~ 30 nm. On the contrary, for ion-irradiated samples (blue), the structural coherence is shortened below the sample thickness (~50 nm) as the dose is increased.

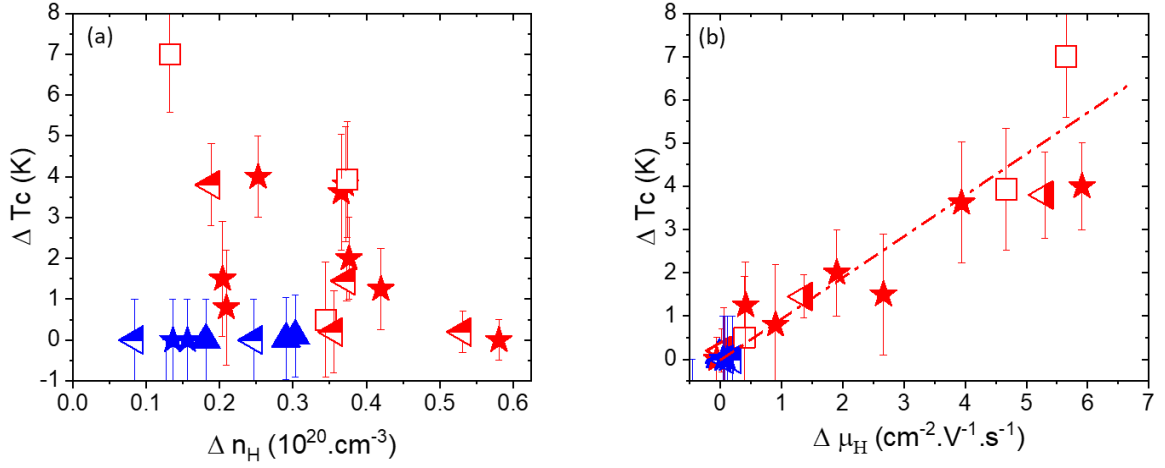

**Fig. S3** Light-induced variations of the critical temperature $\Delta T_c$ as a function of the photoinduced variations of the Hall number (a) and the Hall mobility (b) for the two sets of samples and under all types of illumination. The photoinduced changes in $T_c$ do not correlate with the changes in the Hall number, but scale with the changes in mobility for all the experimental configurations considered. The red line is a guide to the eye.

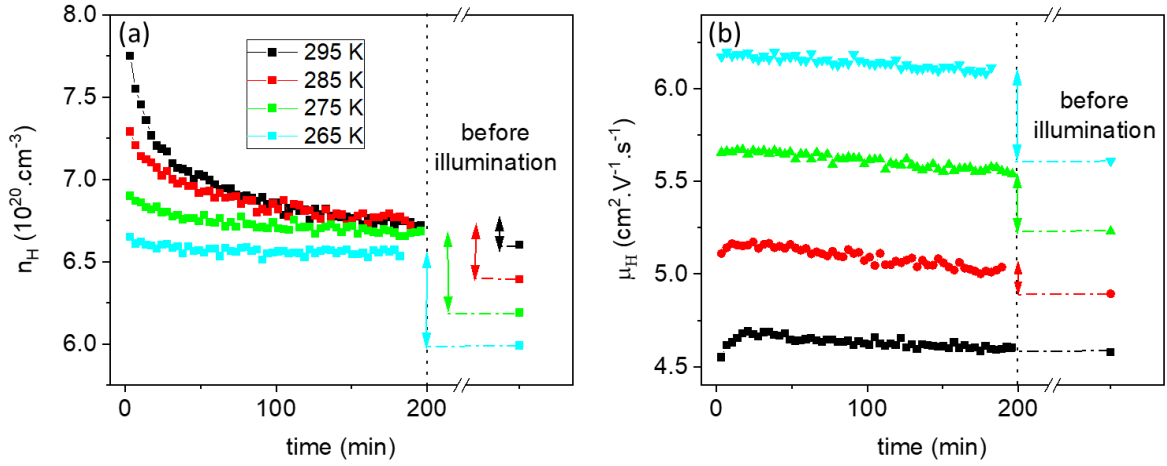

**Fig. S4:** Time-dependent Hall number (a) and Hall mobility (b) measured in the dark at different temperatures (see legend), following illumination at T = 95K with $\lambda = 565\,nm$ and $P = 100\,mW.cm^{-2}$ (for a sample with $x$=6.4). The reference levels on the right were measured before illumination. The double-sided arrows highlight the remnant enhancement at t=200 min. These measurements demonstrate the reproducibility of the results reported in Fig. 3 (main text), which were obtained with another sample of the same nominal oxygen content.

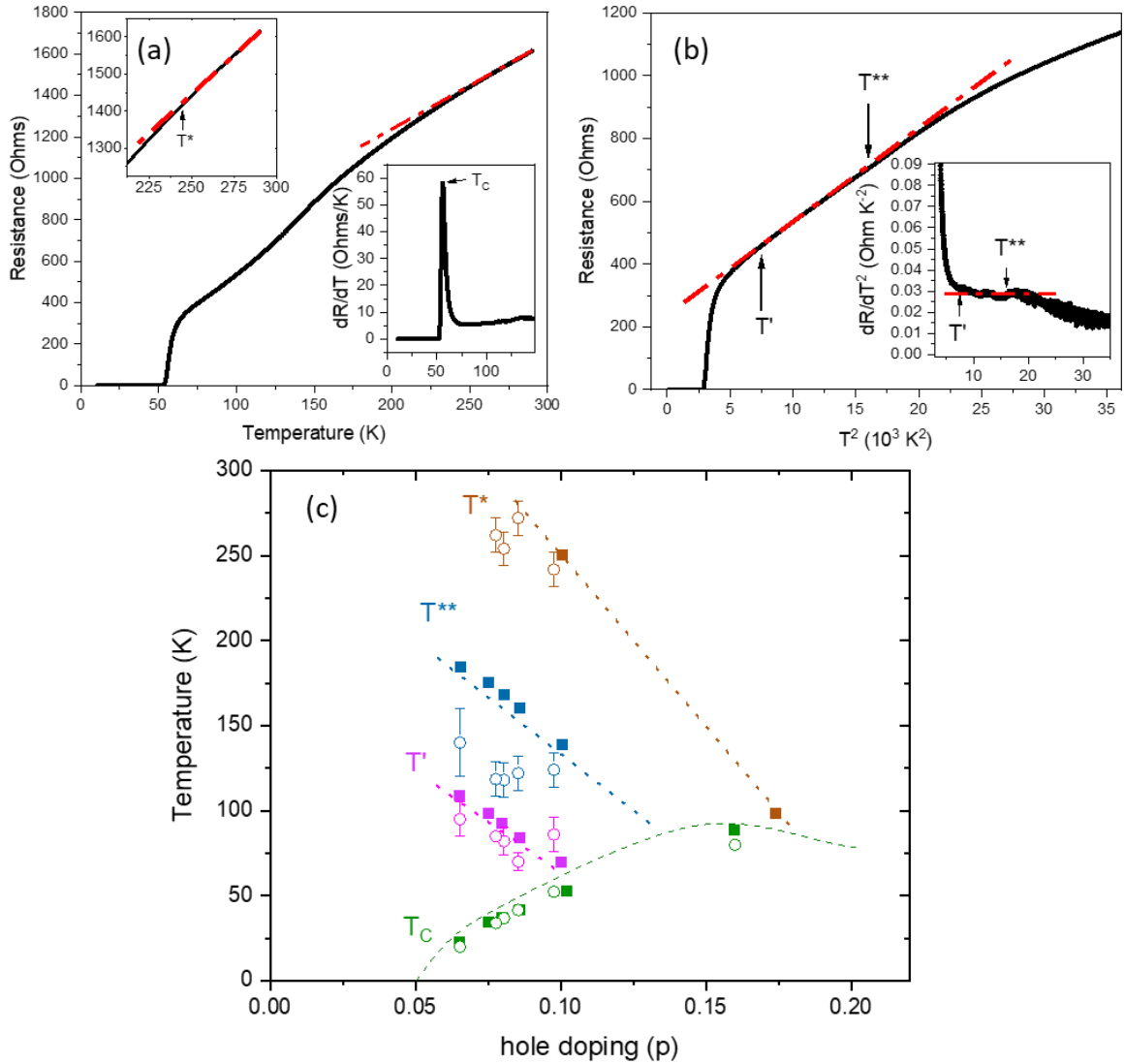

**Figure S5: (a)** R(T) of an oxygen-deficient sample ($x=6.6$). The $T_C$ is taken as the temperature across the transition in which the derivative peaks (lower inset). The R(T) shows the characteristic crossover from the high-temperature linear dependence (strange metal phase) to the quadratic dependence of the pseudogap phase. The departure from linear behaviour occurs at T* (upper inset). **(b)** Resistance *vs*. $T^2$ for the same sample. The crossover into the pseudogap phase leads to a range where the temperature dependence is quadratic, between an upper bond T** and a lower one T' which is usually attributed to the onset of superconducting fluctuations [4]. T** and T' are determined from the derivative $dR/dT^2$ as shown in the inset. **(c)** Phase diagram showing the characteristic temperatures as a function of the hole doping $p$ per Cu. The hole doping $p$ is estimated from the usual [4] phenomenological relationship $1 - T_C/T_C^{max} = 82.6(p - 0.16)^2$, with $T_C^{max}$ the critical temperature for optimum doping. The hollow data points correspond to samples from this work, and the solid data points are taken from the literature [4]. Lines are a guide to the eye. We can see that, despite somewhat lower T**, the studied YBCO films (hollow symbols) present a phase diagram similar that found in the literature (solid symbols).